\documentclass[aps,prl,twocolumn]{revtex4}
\begin{document}
\title{\sc Brownian dynamics approach to interacting magnetic moments}

\author{O.\ Chubykalo}
\affiliation{Instituto de Ciencia de Materiales de Madrid, CSIC,
  Cantoblanco, E-28049 Madrid, Spain}
   \author{R.\ Smirnov-Rueda}
\affiliation{Instituto de Ciencia de Materiales de Madrid, CSIC,
  Cantoblanco, E-28049 Madrid, Spain}
 \author{M.\ A.\ Wongsam}
 \affiliation{Department of Physics, University of Durham, South
  Road, Durham, D1 3LE, UK}
\affiliation{Seagate Research, River Parks Commons, 2403 Sydney Street,
Pittsburgh, PA 15203-2116, USA}
  \author{R.\ W.\ Chantrell}
\affiliation{Department of Physics, University of Durham, South
  Road, Durham, D1 3LE, UK}
\affiliation{Seagate Research, River Parks Commons, 2403 Sydney Street,
Pittsburgh, PA 15203-2116, USA}
\author{U.\ Nowak}
  \affiliation{Theoretische Tieftemperaturphysik,
  Gerhard-Mercator-Universit\"{a}t-Duisburg, 47048 Duisburg, Germany}
  \author{J.M.Gonzalez}
\affiliation{Instituto de Ciencia de Materiales de Madrid, CSIC,
  Cantoblanco, E-28049 Madrid, Spain}

\date{\today}
\begin{abstract}
  The question how to introduce thermal fluctuations in the equation
  of motion of a magnetic system is addressed.  Using the approach of
  the fluctuation-dissipation theorem we calculate the properties of
  the noise for both, the fluctuating field and fluctuating torque
  (force) representation. In contrast to earlier calculations we
  consider the general case of a system of interacting magnetic
  moments without the assumption of axial symmetry.  We show that
  the interactions do not result in any correlations of thermal
  fluctuations in the field representation and that the same widely
  used formula can be used in the most general case. We further prove
  that close to the equilibrium where the fluctuation-dissipation
  theorem is valid, both, field and torque (force) representations
  coincide, being different far away from it.
\end{abstract}
\pacs{} \maketitle


The problem of a correct introduction of temperature in the equation
of motion of a magnetic system has gained much importance as a result
of technological requirements of magnetic recording industry
\cite{Weller,Charap,Bertram}. This is associated with the need to
perform calculations of magnetization dynamics at finite temperatures.
Open problems include fast magnetization switching, thermal stability
and magnetic viscosity, among others. The correct solution of the
problem is still far from being understood.  The main difference
between the magnetic problem and the standard molecular dynamics
approach is that the magnetic moment dynamics is governed by the
Landau-Lifshitz equation which includes the precession of a magnetic
moment around its internal field direction. It comprises coupled
first-order equations for the magnetization components, and the
requirement of conservation of the magnetization amplitude. As a
consequence, no analogue of mass and kinetic energy exist in the
system, thus making it impossible to introduce the temperature through
this mechanism.

Consequently, the temperature is introduced through small deviations
from the equilibrium configuration. Therefore, strictly speaking,
this approach is only valid when these deviations are small and it
cannot be used for fast magnetization switching.

Let us briefly summarize the original approach from W.\ Brown
\cite{Brown,Brown1}. The underlying equation of motion is the
Landau-Lifshitz-Gilbert equation which can be written in the form
\begin{equation}
  \label{9}
  {d \overrightarrow{\textbf{M}_i }\over
    d\tau}=-\overrightarrow{\textbf{M}_i}\times\overrightarrow{\textbf{H}_i}
  -\alpha \overrightarrow{\textbf{M}_i}\times
  [\overrightarrow{\textbf{M}_i}\times \overrightarrow{\textbf{H}_i}],
\end{equation}
where
\begin{equation}
  \label{10}
  \tau= {\gamma_0 H_k \over M_s (1+\alpha^2)} t, \;\;\;\label{intf}
  \overrightarrow{\textbf{H}}=-{\delta E^* \over \delta
    \overrightarrow{\textbf{M}}}
\end{equation}
$\gamma_0$ is the gyromagnetic ratio and $\alpha$ is the damping
constant. The magnetic moment $\overrightarrow{\textbf{M}}$ is
normalized to the saturation value $M_s$, and the internal field
$\overrightarrow{\textbf{H}}$ is normalized to the anisotropy
field $H_k=2K /M_s$. The energy $E^* = E/2KV$, where $K$ is the
anisotropy value and $V$ is the particle volume, contains all the
necessary energy contributions: anisotropy, exchange,
magnetostatic and Zeeman.

W.\ Brown proposed the inclusion of thermal fluctuations via a random
field, added to the internal field, Eq.\ \ref{intf}.  For the
calculation of the properties of the random field he outlined two
methods: (i) based on the fluctuation-dissipation theorem (see also\
\cite{Berkov}) and (ii) by imposing the condition that the equilibrium
solution of the correspondent Fokker-Planck equation is the Boltzmann
distribution (see also \cite{Garcia-Palacios}). As a result of both
the thermal field statistical properties are given by
\begin{equation}
\label{thermf} \langle\xi_i\rangle = 0, \;\;
\langle\xi_i(0)\xi_j(\tau)\rangle={\alpha k_B T \over KV
(1+\alpha^2)} \delta_{ij} \delta(\tau), \end{equation}
 where $i,j$
denote Cartesian components $x,y,z$.  Different approaches based,
for example, on the Landau-Lifshitz rather than on the
Landau-Lifshitz-Gilbert equation were also introduced
\cite{Garcia-Palacios,Moro}.

However, the properties of the thermal noise, Eqs.\ \ref{thermf},
were derived only for one isolated particle. Moreover, Brown
considered in his paper \cite{Brown} only the simplest axially
symmetric case.  Nevertheless, in the past the formulas above
provided the basis for practically every numerical method
\cite{Charap,Lyberatos,Hannay,Nowak} for the computation of
magnetization dynamics taking into account thermal fluctuations.
But the investigated magnetic systems usually comprise interacting
particles \cite{Braun,Nakatani,Boerner,Zhang,Hinzke,Schrefl} due
to magnetostatic and/or exchange couplings. For that case the
thermal field may be expected to be influenced by correlations
between different particles \cite{Hannay}.  Hence, it is necessary
to generalize Brown's result to the case of interacting magnetic
moments. To the best of our knowledge, this has never been done
before.

In what follows we start with the Brownian dynamics approach (see
\cite{LandauLifshitz}) which was originally applied to magnetic
systems by A.\ Lyberatos et al.  \cite{Berkov,Hannay}. However, we
consider the general case of an interacting system with a
non-axially-symmetric potential.  Following the standard approach, we
introduce the temperature into the motion of the Brownian particles
(i.\ e.\ the magnetic moments) as a result of the
fluctuation-dissipation theorem.  Consequently, this approach is only
valid when small deviations from equilibrium are considered.

The general Langevin equation of motion is written in the form
\begin{equation}
  \label{1}
  {dx_i \over dt} = - \sum_j \gamma_{ij} X_j + f_i,\;\;\;
  X_j=-{\partial S \over \partial x_j},
\end{equation}
where the $\gamma_{ij}$ are the so-called kinetic coefficients,
 $X_j$ are variables which are thermodynamically conjugate to
$x_j$, and $S$ is the entropy of the magnetic system. For a closed
system in an external medium,
\begin{equation}
  \label{3}
  X_j={1 \over k_B T} {\partial E \over \partial x_j}.
\end{equation}

In Eq.\ \ref{1}, $f_i$ is a random force representing thermal
fluctuations in the system having the properties
\begin{equation}
  \label{4}
  \langle f_i(t) \rangle = 0 \quad \mbox{and} \quad \langle f_i(0)f_j(t)
  \rangle = \mu_{ij} \delta(t)
\end{equation}
where
\begin{equation}
\label{mu} \mu_{ij} = \gamma_{ij} + \gamma_{ji}.
\end{equation}
A linear equation of motion of the form
\begin{equation}
\label{5} {dx_i \over dt}=\sum_j L_{ij} x_j,
\end{equation}
with the associated energy
\begin{equation}
\label{6}
 E=E_0+{1 \over 2} \sum_{i,j} A_{ij} x_i x_j,
\end{equation}
can be rewritten as
\begin{equation}
\label{7} {dx_i\over dt}=-\sum_j L_{ij}x_j=-\sum_j \gamma_{ij}{1
  \over k_B T} \sum_k A_{kj} x_k,
\end{equation}
so that the matrix $L_{ik}$ is related to the kinetic coefficients
$\gamma_{ik}$ in the following way \cite{Berkov}:
\begin{equation}
  \label{8}
  L_{ik}=-{1 \over k_B T} \sum_j \gamma_{ij} A_{kj}
\end{equation}

In micromagnetics the motion of a magnetic moment $\mathbf M$ is
governed by the deterministic LLG equation (Eq.\ \ref{9}).  For the
equilibrium state of the system Brown's condition
\begin{equation}
  \label{12}
  \overrightarrow{\textbf{M}_i^0} \times
  \overrightarrow{\textbf{H}_i^0}=0
\end{equation}
must be satisfied, implying that here
$\overrightarrow{\textbf{H}_i^0}$ and
$\overrightarrow{\textbf{M}_i^0}$ are parallel. Close to equilibrium,
the LLG equation can be linearized using small deviations
\begin{equation}
  \label{13}
  \overrightarrow{\textbf{m}_i}=\overrightarrow{\textbf{M}_i}-
  \overrightarrow{\textbf{M}_i^0},\,\,\,\,\,
  \overrightarrow{\textbf{h}_i}=\overrightarrow{\textbf{H}_i}-
  \overrightarrow{\textbf{H}_i^0}
\end{equation}
from their equilibrium values, yielding
\begin{equation}
  \label{14}
  \frac{d m_{i}}{dt}=\sum_{j=1}^{3N} L_{ij} m_j.
\end{equation}
Here, the indices $i,j$ count the particles sites $1,...,N$ as
well as their $x,y,z$ coordinates.  The internal fields $h_j$ play
the role of the variables which are thermodynamically conjugate to
$m_j$,
\begin{equation}
  \label{15}
  X_j={1 \over k_BT} {\partial E \over \partial m_j}= -{2KV \over k_B
    T} h_j.
\end{equation}

Thus, the LLG equation should be rewritten in the form
\begin{equation}
  \label{16}
  {dm_i \over dt}={2KV \over k_B T} \sum_{j=1}^{3N} \gamma_{ij}
  h_{j}
\end{equation}
which is an easier way to calculate the kinetic coefficients than the
use of Eq.\ \ref{8}. The representation of the LLG equation in the
form of Eq.\ \ref{1} means that in what follows the thermal
fluctuations are introduced as a fluctuating torque (a generalized
force rather than a field).  Later we will show that in the linear
approximation this is equivalent to the standard fluctuating field
representation. Alternatively, Eq.\ \ref{16} could be viewed as a
polar representation of the magnetization vector $m_{i}^1=\theta_i,
m_{i}^2=\varphi_i$, in this case the conjugate variables are the polar
projections of the internal fields $(h_{\theta},h_{\varphi})$ and the
fluctuations $f_i$ will stand for the random field polar
components. This latter approach was used originally by W.\ Brown
\cite{Brown}.

We continue by writing the energy of the system in the form
\begin{equation}
  \label{17} E^* = \sum_i^N \big( -\overrightarrow{\textbf{M}_i} \cdot
  \overrightarrow{\textbf{H}_i} + {\lambda \over 2} \textbf{M}_i^2
  \big) .
\end{equation}
where $\lambda$ is the Lagrange multiplier. In the zero order
approximation one obtains
\begin{equation}
  \label{18}
  \overrightarrow{\textbf{M}_i^0} = {1 \over \lambda}
  \overrightarrow{\textbf{H}_i^0}
\end{equation}
which corresponds to Brown's condition, Eq.\ \ref{12}. The linear
approximation leads to the equilibrium condition
\begin{equation}
  \label{19}
  -\overrightarrow{\textbf{M}_i^0} \cdot \overrightarrow{\textbf{h}_i} -
  \overrightarrow{\textbf{H}_i^0} \cdot \overrightarrow{\textbf{m}_i} +
  \lambda
  \overrightarrow{\textbf{M}_i^0} \cdot \overrightarrow{\textbf{m}_i} = 0,
\end{equation}
which leaves us only the quadratic form for the energy expression near
the equilibrium,
\begin{equation}
  \label{20}
  E^* = E_0 - \sum_i^N \big( \overrightarrow{\textbf{m}_i} \cdot
  \overrightarrow{\textbf{h}_i} - \frac{\lambda}{2} \textbf{m}_i^2 \big).
\end{equation}

The general expression for magnetic energies is a quadratic form in
terms of the magnetization (apart from the Zeeman term which is
included in the equilibrium field $\overrightarrow{\textbf{H}_i^0}$
and condition \ref{19}). Therefore, it is reasonable to suppose that
the total field can be expressed as
\begin{equation}
  \label{21}
  h_i^{\alpha} = \sum_{j,\beta}B_{ij}^{\alpha \beta}
  m_j^{\beta} = h_{eff,i}^{\alpha}-\lambda m_i^{\alpha}
\end{equation}
where $h_{eff,i}^{\alpha}$ are the components of the effective
field due to the different energy contributions and $-\lambda
m_i^{\alpha}$ is the field due to the kinematic interaction
expressing the constraints (Lagrange multiplier).  The value of
the Lagrange multiplier is normally found from the equilibrium
condition \ref{19}. However, its actual value is not necessary for
calculations due to the fact that the LLG equation conserves the
magnetization length. Latin indices represent the sites of the
moments and the Greek ones the magnetization components $x,y,z$.
In this case the final expression for the energy (Eq.\ \ref{20})
takes the form
\begin{equation}
  \label{22}
  E^* = E_0 -\sum_{i,j,\alpha,\beta}(B_{ij}^{\alpha \beta} -
  \delta_{ij} \delta_{\alpha \beta})m_i^{\alpha} m_j^{\beta}.
\end{equation}

The expressions for the kinetic coefficients could be obtained by
using directly the expression \ref{8}. In this approach it seems that
the final result could also include correlations between different
particles \cite{Hannay}. But this is not the case: the kinetic
coefficients can be obtained much easier representing the linearized
LLG equation in the form of Eq.\ \ref{16} yielding
\begin{eqnarray}
  \gamma_{ij}^{xx}&=&{\alpha k_B T \over 2KV}
  \left[(M_i^{0,y})^2+(M_i^{0,z})^2\right]  \delta_{ij}\\
  \gamma_{ij}^{xy}&=&{k_B T \over 2 KV}\left[-M_i^{0,z}+\alpha
    M_i^{0,x}M_i^{0,y}\right] \delta_{ij}\\
  \gamma_{ij}^{yx}&=&{k_B T \over
    2KV}\left[M_i^{0,z}+\alpha M_i^{0,x}M_i^{0,y}\right] \delta_{ij}\\
  \gamma_{ij}^{yy}&=&{\alpha k_B T \over 2KV}
  \left[(M_i^{0,x})^2+(M_i^{0,z})^2\right] \delta_{ij}.
\end{eqnarray}
The other coefficients can be obtained by symmetry. Note, that
there are no correlations between different particles.  Also, the
kinetic coefficients have obviously reversible parts (coming from
rotation) and irreversible parts (from damping). The reversible
antisymmetric parts do not contribute to the thermal fluctuations
after adding the kinetic coefficients to calculate the matrix
$\mu$ from Eq.\ \ref{mu}, yielding
\begin{eqnarray}
  \label{mu1}
  \mu^{xx}_{ij} & = & {\alpha k_B T \over KV}
  \left[(M_i^{0,y})^2+(M_i^{0,z})^2\right] \delta_{ij}\\
  \label{muxy}
  \mu^{xy}_{ij} & = & {\alpha k_B T \over KV} M_i^{0,x}M_i^{0,y}
  \delta_{ij}.
\end{eqnarray}
Once again, the others can be obtained by symmetry. Note that in a
general system of coordinates there are correlations between different
magnetization components but no correlations between different
particles. However, if we set the local coordinate system such that
the $z$ axis coincides with the equilibrium magnetization direction,
$M_i^{0,x} = 0, M_i^{0,y}=0, M_i^{0,z}=1$, these correlations
disappear and we have the same thermal fluctuations in $x$ and $y$
directions but no fluctuations in $z$ direction,
\begin{equation}
  \label{fluct}
  \mu^{xx}_{ij}=\mu^{yy}_{ij}={ \alpha k_B T \over KV}\delta_{ij}
    \quad \mbox{and}  \quad \mu_{ij}^{zz}=0.
\end{equation}
Thus, the torque fluctuations produce effectively correlations and
different values of thermal fluctuations in all other systems of
coordinates different from the global one, where one of the axes is
parallel to the equilibrium magnetization direction and where the
equation of motion for this component disappears.

It is customary to introduce thermal fluctuations in the field
components (see \cite{Lyberatos} and originally W.\ F.\ Brown
\cite{Brown}) instead of the torque fluctuations as derived above.
This has its origin in the representation of the LLG equation in a
spherical system of coordinates in form of Eq.\ \ref{1}. However, in
both of these papers above only the axially symmetric case without
interactions was considered.  The big difference between these two
approaches is the \emph{multiplicative} character of the field noise
versus the \emph{additive} noise of the torque. This turns out to be
important for larger magnetization deviations. But first we will show
that in the global coordinate system both approaches, torque and
field, give the same result, as long as the magnetization deviations
from the equilibrium are small.

Let us use a decomposition of the field components according to $
H^{i}\rightarrow H^{i}+\xi^{i}$, where $\xi^{i}$ are the components of
the fluctuation part of the field.  When this is done we obtain the
following expansion of the equations of motion,
\begin{eqnarray}
  \nonumber \frac{dM^{i}}{d\tau} & = & -\varepsilon^{ijk}M^{j}H^{k} -\alpha
  H^{m}\left[M^{m}M^{i}-\delta^{mi}\right]\\ \nonumber
  && -\varepsilon^{ijk}M^{j}\xi^{k} -\alpha
  \xi^{m}\left[M^{m}M^{i}-\delta^{mi}\right]\\ \label{eq:langevin}
  & = & A^{i}(M^{n},H^{l})+B^{ij}(M^{n})\xi^{j}.
\end{eqnarray}
Furthermore, in the global system of coordinates we linearize the
magnetization by the decomposition $M^{i}\rightarrow
M_{0}^{i}+m^{i}$, where $m^{i}$ are {\emph small} fluctuations
around the equilibrium values $M_{0}^{i}$, and apply the
constraint condition, $|\overrightarrow{\textbf{M}}| = 1$. For
simplicity below we drop in the formulas the particle index $i$.
The components in the specified coordinate system are then
\begin{eqnarray}
\label{eq:fluc1}
\frac{dm^{x}}{d\tau} & = & A^x(\overrightarrow{m})-\left(m^{y}+\alpha
m^{x}\right)\xi^{z}+f^{x}, \\ \label{eq:fluc2}
\frac{dm^{y}}{d\tau} & = & A^y(\overrightarrow{m})+\left(m^{x}-\alpha
m^{y}\right)\xi^{z}+f^{y}, \\ \label{eq:fluc3}
\frac{dm^{z}}{d\tau} & = &
\left(m^{x}+m^{y}\right)\left(\xi^{x}-\alpha\xi^{y}\right),
\end{eqnarray}
where $\overrightarrow{A}(\overrightarrow{m})$ stands for the
linearized deterministic part of the LLG equation and
\begin{equation}
  \label{ffl}
  f^{x} =  \xi^{y}+\alpha\xi^{x}, \;\;\;
f^{y}  =  -\xi^{y}+\alpha\xi^{x}.
\end{equation}
The constraint condition implies that in a first order
approximation it is $m^{z}(\tau)=0,\,\,\forall\,\tau$. This is
compatible with Eq.\ \ref{eq:fluc3} only if the field fluctuations
$\xi^{i}$ can be considered to be small quantities, in which case
products of the $\xi^{i}$ with the $m^{i}$ can be ignored in
Eq.(\ref{eq:fluc1}). These equations suggest that the field
fluctuations contribute \emph{additively}. From Eqs.\ \ref{ffl}
one can also obtain Brown's formulas for the field fluctuations
(Eq.\ \ref{thermf}).

It is important to note that the last equation is satisfied for any
fluctuation field value due to the character of the
LLG equation. Thus the $f^z$ value (or $\xi^z$) is
in this case undefined. In any case the component $\xi^z$ is not
efficient since it acts parallel to the magnetization direction. The
assumption made in the paper of A.\ Lyberatos and R.\ Chantrell
\cite{Lyberatos} is that the field components are isotropic and that
\begin{equation}
\label{filed}
 \langle\xi^x\rangle=\langle\xi^y\rangle=\langle\xi^z\rangle.
\end{equation}
This assumption in the global system of coordinates (where the
fluctuation-dissipation theorem is applied) leads to the remarkable
symmetry (\ref{filed}) of the field components in all the systems of
coordinates and to the absence of correlations.  Furthermore, it is
assumed that this property is valid through the magnetization
reversal.

For the torque fluctuations the reasonable hypothesis to mimic the
field ones would be the assumption that there are never torque
(force) fluctuations along the magnetization direction. In this
case the correlations between different noise components would
appear in all other systems of coordinates different from the
global one. While equivalent near the equilibrium, these two
approaches will be different far from it. At this point, we would
like to restate that the whole theory is valid for small
fluctuations around the equilibrium where both approaches
coincide.

In conclusion, the application of the Brownian dynamics approach to
the motion of a magnetic system shows that interactions do not
introduce correlations into thermal fluctuations introduced as both,
either a fluctuating torque or a fluctuating field. Correlations may
appear between different magnetization components as a result of the
conservation of the value of the magnetic moment. The reasonable
hypothesis that all the fluctuating field components are equivalent
leads to Brown's well-known formulas for the fluctuating fields values
without correlations. This validates all previously done micromagnetic
calculations where this kind of assumption was made.

\begin{acknowledgments}
  OC acknowledges the hospitality and support from Durham University,
  UK, Duisburg University, Germany, and Seagate Research Center,
  Pittsburg, USA, where a part of this work was done. RS-R and UN
  thank Durham University, UK, for hospitality and support. MAW thanks
  ICMM, Madrid, Spain for hospitality and support, and acknowledges
  the EPSRC who supported this project under grant No.  R040 318. The
  authors acknowledge useful discussions with A.\ Lyberatos (Seagate
  Research Center, Pittsburgh, USA).
\end{acknowledgments}

%


\begin{thebibliography}{99}
\bibitem{Weller} D. Weller and A. Moser, IEEE Trans. Magn. {\bf 35},
  4425, (1999).

\bibitem{Charap} Y. Kanai and S. H. Charap, IEEE Trans. Magn. {\bf 27},
  4972, (1991).

\bibitem{Bertram} Y. Zhang and H. N. Bertram, IEEE Trans. Magn. {\bf
    34}, 3786 (1998).

\bibitem{Brown} W. F. Brown, Phys. Rev. {\bf 130}(5), 1677 (1963)

\bibitem{Brown1} W. F. Brown, IEEE Trans. Magn. {\bf MAG-15}  (1979).

\bibitem{Berkov} A. Lyberatos, D.~V. Berkov, and R.~W. Chantrell,
  J.~Phys.: Condens.~Matter {\bf 5}, 8911 (1993).

\bibitem{Garcia-Palacios} J. L. Garcia-Palacios and F.-J. L\'{a}zaro,
  Phys. Rev. B {\bf 58}, 14937 (1998).

\bibitem{Moro} T. Kamppeter, F. Mertens, E. Moro, A. S\'{a}nchez and
A. R. Bishop, Phys. Rev. B {\bf 59}, 11439 (1999).

\bibitem{Lyberatos} A. Lyberatos, R. W. Chantrell, J.  Appl. Phys. {\bf
    73}(10), 6501 (1993).

\bibitem{Hannay} R. W. Chantrell, J. D. Hannay, M. Wongsam, T. Schrefl
  and H. J. Richter, IEEE Trans.\ Magn.\ {\bf 34}, 1839 (1998).

\bibitem{Nowak} U. Nowak, R. W. Chantrell, and E. C. Kennedy, Phys. Rev.
  Lett. {\bf 84}(1), 163 (2000).

\bibitem{Braun} H.~B. Braun, Phys.\ Rev.\ Lett. {\bf 71}, 3557 (1993).

\bibitem{Nakatani} Y.\ Nakatani and Y.\ Uesaka and N.\ Hayashi and H.\
  Fukushima, J.\ Magn.\ Magn.\ Mat. {\bf 168}, 347 (1997).

\bibitem{Boerner} E.~D. Boerner and H.~N. Bertram, IEEE Trans.\ Mag.
  {\bf 33}, 3052 (1997).

\bibitem{Zhang} K. Zhang and D.~R. Fredkin, J.\ Appl.\ Phys. {\bf 85},
  5208 (1999).

\bibitem{Hinzke} D. Hinzke and U. Nowak, Phys.\ Rev.\ B {\bf 61}, 6734
  (2000) and J.\ Magn.\ Magn.\ Mat. {\bf 221}, 365 (2000).

\bibitem{Schrefl} W.\ Scholz, T.\ Schrefl and J.\ Fidler, J.\ Magn.\
  Magn.\ Mat. {\bf 233}, 296 (2001).

\bibitem{LandauLifshitz} L.\ D.\ Landau and E.\ M.\ Lifshitz,
Course of Theoretical Physics V: Statistical Physics, Pergamon
Press, Oxford 1969

\end{thebibliography}
\end{document}